\documentclass[twocolumn,prd,superscriptaddress,showpacs,floatfix,%
preprintnumbers,nofootinbib]{revtex4}  
  
\usepackage{epsfig}  

\long\def\ignore#1{}  
  
\usepackage[dvips]{color}  
\definecolor{Black}{named}{Black}  
\definecolor{Blue}{named}{Blue}  
\definecolor{Red}{named}{Red}

\newcommand{\D}{{\rm d}}

\newcommand{\wsH}{\omega^{\rm s}_{\rm H}}  
\newcommand{\wsL}{\omega^{\rm s}_{\rm L}}  
  
\begin{document}  
  
\title{Spectral split in prompt supernova neutrino burst:  
Analytic three-flavor treatment}  
  
\author{Basudeb Dasgupta}  
\affiliation{Tata Institute of Fundamental Research, Homi Bhabha  
Road, Mumbai 400005, India}  
  
\author{Amol Dighe}  
\affiliation{Tata Institute of Fundamental Research, Homi Bhabha  
Road, Mumbai 400005, India}  
  
\author{Alessandro Mirizzi}  
\affiliation{Max-Planck-Institut f\"ur Physik  
(Werner-Heisenberg-Institut), F\"ohringer Ring 6, 80805 M\"unchen,  
Germany}  
  
\author{Georg G.~Raffelt}  
\affiliation{Max-Planck-Institut f\"ur Physik  
(Werner-Heisenberg-Institut), F\"ohringer Ring 6, 80805 M\"unchen,  
Germany}  
  
\date{10 January 2008, revised 28 April 2008}  
  
\preprint{MPP-2008-3}  
  
\begin{abstract}  
The prompt $\nu_e$ burst from a core-collapse supernova (SN) is  
subject to both matter-induced flavor conversions and strong  
neutrino-neutrino refractive effects. For the lowest-mass  
progenitors, leading to O-Ne-Mg core SNe, the matter density profile  
can be so steep that the usual MSW matter effects occur within the  
dense-neutrino region close to the neutrino sphere. In this case a  
``split'' occurs in the emerging spectrum, i.e., the $\nu_e$ flavor  
survival probability shows a step-like feature. We explain this  
feature analytically as a ``MSW prepared spectral split.'' In a  
three-flavor treatment, the step-like feature actually consists of  
two narrowly spaced splits. They are determined by two combinations  
of flavor-lepton numbers that are conserved under collective  
oscillations.  
\end{abstract}  
  
  
\maketitle  
  
\section{Introduction}                        \label{sec:introduction}  
  
The neutrino flux streaming off a collapsed supernova (SN) core is an  
intriguing astrophysical case where flavor transformations depend  
sensitively on some of the unknown elements of the leptonic mixing  
matrix~\cite{Dighe:1999bi,Dighe:2007ks}. Since the conversion  
probabilities depend also on the time-dependent matter profile, a  
high-statistics SN neutrino observation may also reveal, for example,  
signatures for shock-wave propagation~\cite{Schi, Foglish, Tomas,  
FogliMega, Dasgupta:2005wn, Fogli:2006xy, Friedland:2006ta}. While  
galactic SNe are rare, various ongoing and future experimental  
programmes depend on large detectors that, besides their main  
purpose, are also sensitive to SN neutrinos~\cite{Autiero:2007zj}.  
Therefore, understanding the flavor evolution of a SN neutrino signal  
remains of topical interest.  
  
The flavor transformation probabilities not only depend on the matter  
background, but also on the neutrino fluxes themselves:  
neutrino-neutrino interactions provide a nonlinear term in the  
equations of motion~\cite{Pantaleone:1992eq, Sigl:1992fn} that causes  
collective flavor transformations~\cite{Samuel:1993uw,  
  Kostelecky:1993dm, Kostelecky:1995dt, Samuel:1996ri, Pastor:2001iu,  
  Wong:2002fa, Abazajian:2002qx, Pastor:2002we, Sawyer:2004ai, Sawyer:2005jk}. Only  
recently has it been fully appreciated that in the SN context these  
collective effects give rise to qualitatively new  
phenomena~\cite{Duan:2005cp, Duan:2006an, Hannestad:2006nj,  
  Duan:2007mv, Raffelt:2007yz, EstebanPretel:2007ec, Raffelt:2007cb,  
  Raffelt:2007xt, Duan:2007fw, Fogli:2007bk, Duan:2007bt, Duan:2007sh,  
  EstebanPretel:2007yq, Dasgupta07}.  
  
One peculiar aspect of the expected SN neutrino fluxes is the  
hierarchy $F_{\nu_e} > F_{\bar\nu_e} > F_{\nu_\mu} = 
F_{\bar\nu_\mu} =F_{\nu_\tau} = F_{\bar\nu_\tau}$  
so that there is an excess flux of  
$\nu_e\bar\nu_e$ pairs over those of the other flavors. 
The nonlinear terms cause a collective transformation  
$\nu_e\bar\nu_e\to\nu_x\bar\nu_x$,
where $\nu_x$ is a specific linear combination of 
$\nu_\mu$ and $\nu_\tau$.
 The detailed dynamics of this  
transition is complicated and several important aspects are only  
numerically observed, not analytically understood. Still, the most  
crucial point is that the pair transformation  
$\nu_e\bar\nu_e\to\nu_x\bar\nu_x$ proceeds collectively much faster  
than ordinary pair annihilation, so we have to contend with a ``speed-up  
phenomenon'' \cite{Sawyer:2004ai, Sawyer:2005jk}. The pair process  
does not violate flavor-lepton number. Being an instability in flavor  
space, it proceeds efficiently even for a very small mixing angle.  
  
\begin{figure}[b]  
\includegraphics[angle=0,width=0.8\columnwidth]{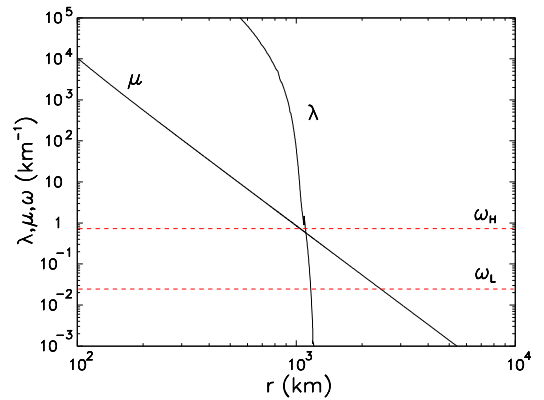}  
\caption{Profile of the matter potential $\lambda$ and the effective  
neutrino-neutrino interaction potential $\mu$ for an O-Ne-Mg core  
collapse SN~\cite{Nomoto1984, Nomoto1987,  
Kitaura:2005bt,Janka:2007di}.\label{fig:profile}}  
\end{figure}  
  
\begin{figure*}  
\includegraphics[angle=0,width=0.6\textwidth]{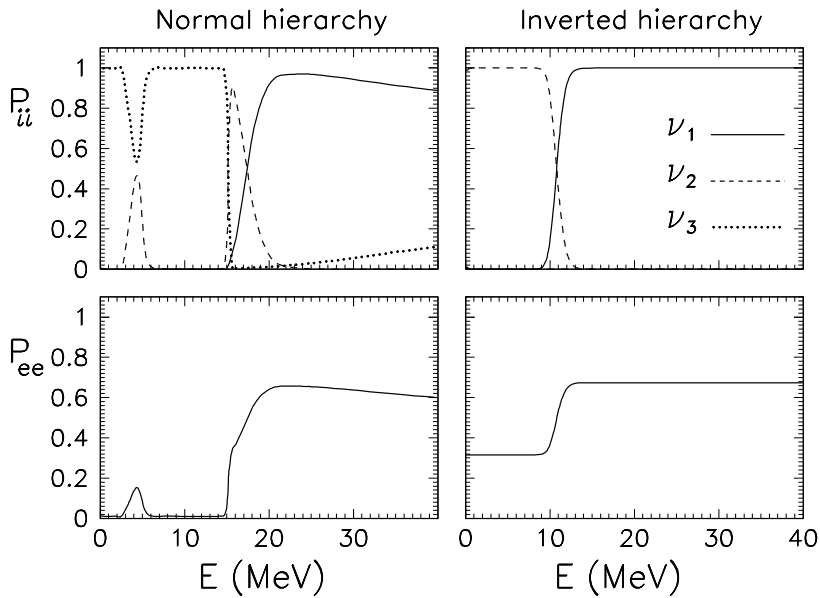}  
\caption{Mass eigenstate fractions $P_{ii}$ 
as well as the $\nu_e$ survival probabilities far away from 
the star, 
numerically computed using the SN model of Fig.~\ref{fig:profile}
and an initial flux of pure $\nu_e$.
\label{fig:numericalsplit}}  
\end{figure*}  
  
A very different situation prevails in the interior of a SN core  
where the $\nu_e$ distribution is determined by a large chemical  
potential that enhances the $\nu_e$ density and suppresses the  
$\bar\nu_e$ density relative to that of $\nu_x$ and $\bar\nu_x$ so  
that collective pair conversions are not possible. Neutrino-neutrino  
interactions are strong, but their only impact is to synchronize the  
flavor oscillations (``self-maintained coherence''). Significant  
flavor transformation here requires a violation of flavor-lepton  
number. However, the high density of ordinary matter suppresses the  
effective mixing angles between $\nu_e$ and the other flavors.  
Therefore, in the interior of a SN core the individual flavor-lepton  
numbers are almost perfectly conserved~\cite{Hannestad:1999zy}.  
  
Immediately after collapse, the SN emits a prompt $\nu_e$ burst that  
arises from the de-leptonization (neutronization) of the outer layers  
of the collapsed core. Once more we have a strongly enhanced $\nu_e$  
and a suppressed $\bar\nu_e$ flux relative to the other  
flavors~\cite{Kachelriess:2004ds}. Once more, collective pair  
transformations are not possible: efficient flavor conversion  
requires a large violation of flavor-lepton number and is not  
possible if the ordinary matter density is large. At some distance  
from the neutrino sphere, the $\nu_e$ flux encounters the usual MSW  
level crossings driven by the atmospheric neutrino mass difference  
$\Delta m^2_{\rm atm}$ (H~crossing) and by the solar mass difference  
$\Delta m^2_{\odot}$ (L~crossing), leading to the usual resonant  
transformations~\cite{Dighe:1999bi,Kachelriess:2004ds}.  
  
An interesting new case is motivated by the insight that SNe with the  
lowest progenitor masses of 8--10~$M_\odot$, encompassing perhaps 30\%  
of all cases, collapse before forming an iron core, the class of  
O-Ne-Mg core collapse SNe~\cite{Nomoto1984, Nomoto1987,  
  Kitaura:2005bt, Janka:2007di}. In state-of-the-art numerical  
simulations these SNe explode even in a spherically symmetric  
treatment (convection plays no role), largely because their envelope  
mass is very small. By the same token, the matter density profile  
above the core is very steep even at the time of core bounce.  In  
this case the H and L level crossings occur very close to the  
neutrino sphere and may well lie deeply within the collective  
neutrino region. This is illustrated in Fig.~\ref{fig:profile} where  
we show $\lambda(r)=\sqrt{2}G_{\rm F}n_e(r)$ of an O-Ne-Mg core  
progenitor star~\cite{Nomoto1984, Nomoto1987, Kitaura:2005bt}. We  
also show $\omega_{\rm H}=\langle\Delta m^2_{\rm atm}/2E\rangle$ and  
$\omega_{\rm L}=\langle\Delta m^2_{\odot}/2E\rangle$ as horizontal  
lines, where the average is over the Fermi-Dirac spectrum of neutrino  
energies described below.  The intersection of $\lambda(r)$ with  
these lines indicates the locations of the H and L level crossings.  
  
In Fig.~\ref{fig:profile} we also show the effective  
neutrino-neutrino interaction potential $\mu=\sqrt{2}G_{\rm  
F}F_{\nu_e}\langle 1-\cos\theta\rangle_{\rm eff}$, where $\theta$ is  
the angle between different neutrino trajectories and  
$\langle\ldots\rangle_{\rm eff}$ stands for a suitable average.  At  
large distances, $\mu$~scales approximately as $r^{-4}$.  Collective  
neutrino effects driven by $\Delta m^2_{\rm atm}$ are important for  
$\mu(r)\agt\omega_{\rm H}$ and driven by $\Delta m^2_{\odot}$ for  
$\mu(r)\agt\omega_{\rm L}$.

Duan et al.~\cite{Duan:2007sh} have recently shown that in this case  
the interplay of ordinary MSW conversions with collective oscillations  
leads to interesting effects. 
We start with a pure $\nu_e$ flux with a Fermi-Dirac spectrum ($\langle  
E_{\nu_e}\rangle=11$~MeV, degeneracy parameter $\eta=3$), and
numerically calculate the mass eigenstate fractions $P_{ii}$ and the
$\nu_e$ survival probabilities $P_{ee}$ far away from the star, 
as shown in Fig.~\ref{fig:numericalsplit}.
These plots are in qualitative agreement with the corresponding curves 
in Fig.~2 of  Ref.~\cite{Duan:2007sh}. 
 However, 
our $P_{ee}$ is constructed as an incoherent sum of the mass fractions, 
thus representing the physical situation far away from the star, where
the oscillatory features seen in Duan et al.'s $P_{ee}$ have disappeared.

In inverted hierarchy, one observes that
the neutrinos emerging from the star are in the $\nu_2$ state
at low energies and in the $\nu_1$ state at high energies,
with the transition taking place around $E \approx 12$ MeV.
This results in a step function in energy for $P_{ee}$.

In the normal hierarchy, 
the neutrinos emerging from the star are in $\nu_1$ state
for $E \gtrsim 17$ MeV, in $\nu_2$ state for $15 \, {\rm MeV}
\; \lesssim E \lesssim 17$ MeV, and in the $\nu_3$ state for
$E \lesssim 15$ MeV. The bump seen around $5$ MeV is due to an abrupt change 
in the matter density profile used for the computation 
(see \cite{Duan:2007sh} for details), 
and we do not address it here. 
The transition at $E\approx 15$ MeV is rather sharp, however
the one at $E \approx 17$ MeV is not as abrupt.
This results in a two-step function for $P_{ee}$, 
with the step at $E \approx 17$ MeV somewhat smoothened out.

Broadly, this is an example of a ``MSW prepared spectral split.'' In a  
two-flavor language, it is explained as follows. The strong  
neutrino-neutrino interactions lead to a synchronization of the  
neutrino oscillations. The flavor polarization vector of the ensemble  
begins at high density essentially aligned with the $\nu_e$ direction  
in flavor space. After passing the MSW region, the polarization vector  
emerges with a significant transverse component relative to the mass  
direction because the MSW transition is not fully adiabatic.  
Subsequently this MSW-prepared initial condition is subject to  
collective effects only. As the effective neutrino-neutrino  
interaction becomes weaker, the modes above a certain energy $E_{\rm  
  split}$ orient themselves along the mass direction, those with  
smaller energies in the opposite direction. This is precisely the  
``neutrino only'' case studied in Refs.~\cite{Raffelt:2007cb,  
  Raffelt:2007xt} as a generic case for a spectral split, a case that  
requires one to prepare the polarization vector with a large  
transverse component. (For neutrinos plus antineutrinos, the MSW  
preparation is not necessary because the collective pair  
transformations alone engineer a split.)  
  
Our main goal here is to use the picture of a MSW-prepared spectral  
split to derive analytically the main features seen in Duan et al.'s  
numerical study~\cite{Duan:2007sh},
viz. the existence and the positions of the spectral splits.

The numerical model in Fig.~\ref{fig:profile} shows that the  
MSW resonances are within the collective neutrino region, but not  
deeply inside.  Therefore it is not a priori obvious if the 
``synchronized MSW preparation'' and the subsequent ``split'' 
can be clearly separated.
For our discussion we therefore adopt a more 
schematic model. We artificially increase  
the neutrino-neutrino interaction strength (raise the  
$\mu$-profile in Fig.~\ref{fig:profile}) such that the MSW region and  
the spectral split regions are clearly separate. In this framework we  
first calculate the MSW preparation analytically and then study a  
three-flavor treatment of the spectral split, based on the machinery  
recently developed by two of us~\cite{Dasgupta07}. We find that our  
analytic treatment reproduces the numerical results surprisingly  
well. It also reproduces all relevant features of the realistic 
case, i.e. with the $\mu$-profile of Fig.~\ref{fig:profile}, justifying our 
simplifying assumptions and verifying our general interpretation.  
  
We first set up in Sec.~\ref{sec:SNmodel} our schematic SN  
model that captures the features relevant for our treatment and  
continue in Sec.~\ref{sec:eoms} with the equations of motion. In  
Sec.~\ref{sec:MSW} we derive analytically the MSW-prepared  
three-flavor state that serves as input for our three-flavor spectral  
split study in Sec.~\ref{sec:splitting}. We conclude in  
Sec.~\ref{sec:conclusions}.  
  
\section{Simplified supernova model}               \label{sec:SNmodel}  

\subsection{Realistic scenario}

We take the electron density profile $n_e$ obtained   
from numerical studies of O-Ne-Mg core SNe~\cite{Nomoto1984,  
Nomoto1987, Kitaura:2005bt}, providing $\lambda(r)$ as shown in our  
Fig.~\ref{fig:profile}. The neutrino luminosity is assumed to be  
$L_{\nu_e}=10^{53}~{\rm  
  erg}~{\rm s}^{-1}$ with a Fermi-Dirac spectrum with the average  
energy $\langle E_{\nu_e}\rangle=11$~MeV and a degeneracy parameter  
$\eta=3$, implying a temperature $T=2.76$~MeV. The neutrino sphere is  
taken at the radius $R=60$~km, implying an effective neutrino-neutrino  
interaction strength at large distances~of  
\begin{equation}\label{eq:muprofile}  
\mu(r)=\mu_0\,\left(\frac{{\rm km}}{r}\right)^4  
\end{equation}  
with $\mu_0 = 8.6 \times 10^{11}~{\rm km}^{-1}$. 
This is the profile shown in Fig.~\ref{fig:profile}. 
Based on this model we have solved  
the equations of motion numerically and found the electron neutrino  
survival probability shown in Fig.~\ref{fig:numericalsplit}, in  
agreement with the results of Duan et al.~\cite{Duan:2007sh}.  

In this realistic situation, the decrease of the effective 
neutrino-neutrino interaction with radius is such that the 
spectral split is essentially adiabatic.
Quantitatively, the length scale 
$\ell_\mu \equiv |d \ln\mu(r)/dr|^{-1}$ is 
large enough to satisfy the adiabaticity condition
(see Sec. V of ~\cite{Raffelt:2007cb}) 
during the spectral split.
The very development of a step-like feature in $P_{ee}$
in Fig.~\ref{fig:numericalsplit} is an expression of the adiabaticity.
The ``sharpness'' of the step in $P_{ee}$
is a measure of the degree of adiabaticity, an extremely slowly decreasing  
neutrino-neutrino interaction corresponds to a perfect step  
function in $P_{ee}$~\cite{Raffelt:2007cb,Raffelt:2007xt}. 

\subsection{Analytical treatment}
 
Our analytic treatment is based on a schematic representation of the 
essentials of this realistic case.  
We picture that the synchronized MSW effect occurs first and factorizes from the 
collective oscillations that lead to the spectral split.
To this end, we consider the limiting case where 
$\mu_0$ in eq.~(\ref{eq:muprofile}) is arbitrarily large.
This makes the H and L level crossings more synchronized, 
and pushes the regions where the spectral splits occur, i.e. 
the regions where $\mu(r)$ becomes comparable to 
$\omega_{\rm H}$ or $\omega_{\rm L}$, to much larger radii.
This also renders the spectral splits even 
more adiabatic \footnote{Since $\mu(r)$ is a power law, the length scale
$\ell_\mu$ increases with increasing $r$. As a result,
larger radius for spectral split implies larger $\ell_\mu$
and more adiabaticity}, making the steps in $P_{ee}$ sharper.

It is also important that the MSW transition is not perfectly  
adiabatic. In our schematic model, we use the power-law profile 
for the matter potential,  
\begin{equation}\label{eq:SNprofile2}  
\lambda(r)=\lambda_0\,\left(\frac{r_0}{r}\right)^a  
\end{equation}  
with $\lambda_0=10^3~{\rm km}^{-1}$, $r_0=900$~km and $a=50$. 
This  approximates reasonably the numerical $\lambda(r)$ profile of  
Fig.~\ref{fig:profile} in the neighborhood of the H and L crossings.  
This allows an analytic estimate of the level crossing probabilities 
at the two resonances.

Thus, we obtain our analytic results in the limit of 
$\mu_0\to\infty$ and a power law profile for $\lambda(r)$, where
(i)~the MSW  transitions are perfectly synchronized but semiadiabatic, 
(ii)~the spectral split regions are well separated from the MSW region, and
(iii)~the spectral splits are perfectly adiabatic, making 
the steps in $P_{ee}$  infinitely sharp.
Apart from these minor changes, the physical pictures with the realistic
profile and our schematic profile are identical.

\subsection{Numerical treatment}

We also study the spectral  
splits numerically, assuming the analytic MSW-prepared spectra as  
input. For such a numerical illustration we employ
\begin{equation}\label{eq:mucoefficient}  
\mu_0 = 10^{15}~{\rm km}^{-1}\,,  
\end{equation}  
much larger than the realistic value shown in 
Fig.~\ref{fig:profile}.  

The analytic results match the numerical ones in all
details, and reproduce the main features of the 
realistic situation, with a minor difference that the
spectral splits appear sharper. This is because the net 
effect of our approximations is only to make the MSW 
resonances more synchronized, and the spectral splits 
more adiabatic.

\section{Equations of motion}                         \label{sec:eoms}  
  
\subsection{Matrices of density}  
  
Mixed neutrinos are described by matrices of density $\varrho_{\bf p}$  
for each momentum. The diagonal entries are the usual occupation  
numbers whereas the off-diagonal terms encode phase information. The  
equations of motion (EOMs) are  
\begin{equation}\label{eq:eoms}  
{\rm i} d_t\varrho_{\bf p}=[{\sf H}_{\bf p},\varrho_{\bf p}]\,,  
\end{equation}  
where the Hamiltonian is~\cite{Sigl:1992fn}  
\begin{equation}\label{eq:hamiltonian}  
 {\sf H}_{\bf p}=\Omega_{\bf p}  
 +{\sf V}+\sqrt{2}\,G_{\rm F}\!  
 \int\!\frac{\D^3{\bf q}}{(2\pi)^3}  
 \left(\varrho_{\bf q}-\bar\varrho_{\bf q}\right)  
 (1-{\bf v}_{\bf q}\cdot{\bf v}_{\bf p})\,,  
\end{equation}  
${\bf v}_{\bf p}$ being the velocity. The matrix of vacuum oscillation  
frequencies is $\Omega_{\bf p}={\rm diag}(m_1^2,m_2^2,m_3^2)/2|{\bf  
  p}|$ in the mass basis. The matter effect is represented, in the  
weak interaction basis, by ${\sf V}=\sqrt{2}\,G_{\rm F}n_e\,{\rm  
  diag}(1,0,0)$. While in general there is a second-order difference  
between the $\nu_\mu$ and $\nu_\tau$ refractive  
index~\cite{Botella:1986wy} that can be important for collective  
neutrino oscillations~\cite{EstebanPretel:2007yq}, for the low matter  
densities relevant in our case this ``mu-tau matter term'' is  
irrelevant.  
  
The factor $(1-{\bf v}_{\bf q}\cdot{\bf v}_{\bf p})$ in ${\sf H}_{\bf  
p}$ implies ``multi-angle effects'' for neutrinos moving on different  
trajectories \cite{Sawyer:2004ai, Sawyer:2005jk, Duan:2006an}.  
However, for realistic SN conditions the modifications are small,  
allowing for a single-angle approximation~\cite{Duan:2006an,  
EstebanPretel:2007ec}. In the strongly synchronized regime this is  
not surprising: the strong neutrino-neutrino interaction causes  
self-maintained coherence not only between different energy modes,  
but also between different angular modes. It has not been explained  
why the single-angle approximation remains good even when the  
neutrino-neutrino interaction becomes weak, although numerically this  
is observed to be the case~\cite{Duan:2006an, EstebanPretel:2007ec,  
Fogli:2007bk}.  
  
We are studying the spatial evolution of the neutrino fluxes in  
a quasi-stationary situation. Therefore, the matrices $\varrho_{\bf  
  p}$ do not depend on time explicitly so that the total  
time derivative in the EOMs reduces to the Liouville term involving  
only spatial derivatives. (See Ref.~\cite{Cardall:2007zw} for a  
recent comprehensive discussion of the role of the Liouville term in  
Boltzmann collision equations involving oscillating neutrinos.)  
Moreover, we consider a spherically symmetric system so that the only  
spatial variable is the radial coordinate $r$. In the single-angle  
approximation we finally need to study the simple EOMs,  
\begin{equation}\label{eq:eoms2}  
{\rm i}\partial_r\varrho_{\omega}=[{\sf H}_{\omega},\varrho_{\omega}]\,,  
\end{equation}  
where we now classify different modes by the variable  
\begin{equation}  
\omega=\frac{\Delta m^2_{\rm atm}}{2E}\,.  
\end{equation}  
The single-mode Hamiltonians are  
\begin{equation}\label{eq:hamiltonian2}  
 {\sf H}_{\omega}=\Omega_{\omega}  
 +\lambda(r){\sf L}  
 +\mu(r)\,\varrho\,.  
\end{equation}  
Here we have introduced $\lambda(r)=\sqrt{2}G_{\rm F}n_e(r)$ and  
${\sf L}={\rm diag}(1,0,0)$ in the weak interaction basis. The matrix  
of the total density $\varrho=\int d\omega\varrho_\omega$ is  
normalized such that at the neutrino sphere it is $\varrho={\rm  
diag}(1,0,0)$ in the weak interaction basis. It is conserved except  
for oscillation effects, i.e., the physical neutrino density has been  
absorbed in the coefficient $\mu(r)$ that measures a suitable angular  
average of the neutrino-neutrino interaction energy. The radial  
variation is $\mu(r)\propto r^{-4}$ where a factor $r^{-2}$ comes  
from the geometric flux dilution, another approximate factor $r^{-2}$  
from the fact that neutrinos become more collinear with distance from  
the source.  
  
\subsection{Mixing parameters}  
  
Since ${\sf H}_\omega$ appears in a commutator we may arbitrarily add  
terms proportional to the 3$\times$3 unit matrix. It will prove  
convenient to express the matrix of vacuum oscillation frequencies in  
the form  
\begin{equation}  
\Omega_\omega=\omega\,{\rm diag}\left(-{\textstyle\frac{1}{2}}\alpha,  
+{\textstyle\frac{1}{2}}\alpha,\pm 1\right)\,,  
\end{equation}  
where the mass hierarchy parameter is  
\begin{equation}  
\alpha\equiv  
\frac{\Delta m^2_\odot}{\Delta m^2_{\rm atm}}  
\approx\frac{1}{30}\,.  
\end{equation}  
A positive sign in the third component of $\Omega_\omega$ signifies  
the normal mass hierarchy, a negative sign the inverted hierarchy.  
For the mass differences themselves we  
use~\cite{GonzalezGarcia:2007ib}  
\begin{eqnarray}  
\Delta m^2_{\rm atm}&=&2.4\times10^{-3}~{\rm eV}^2\,,  
\nonumber\\*  
\Delta m^2_{\odot}&=&8\times10^{-5}~{\rm eV}^2\,.  
\end{eqnarray}  
For the mixing angles we use  
\begin{eqnarray}  
\theta_{12}&=&0.6\,,  
\nonumber\\*  
\theta_{23}&=&\pi/4\,,  
\nonumber\\*  
\theta_{13}&=&0.1\,.  
\end{eqnarray}  
Of course, for $\theta_{13}$ only upper limits exist.  
In this context the CP phase $\delta_{\rm CP}$ can be  
ignored, since it does not influence the relevant probabilities  
for equal $\nu_\mu$ and $\nu_\tau$  
fluxes~\cite{Akhmedov:2002zj,Balantekin:2007es}.  
  
\subsection{Bloch vectors}  
  
In the two-flavor context it is well known that the matrices of  
density can be expressed in terms of Bloch vectors, leading to EOMs  
that resemble the precession of a gyroscope around an external force  
field. This picture helps to recognize properties of the EOMs that  
are difficult to fathom in the commutator form of the EOMs.  
Therefore, we follow a recent paper by two of us~\cite{Dasgupta07}  
and note that every Hermitean 3$\times$3 matrix ${\sf X}$ can be  
expressed in the form  
\begin{equation}  
{\sf X}=\frac{1}{3}\,X_0  
+\frac{1}{2}\,{\bf X}\cdot {\bf\Lambda}\,,  
\end{equation}  
where $X_0={\rm Tr}({\sf X})$, ${\bf X}$ is an eight-dimensional  
Bloch vector, and ${\bf\Lambda}$ is a vector of the Gell-Mann  
matrices. Note that $\Lambda_3={\rm diag}(1,-1,0)$ and  
$\Lambda_8={\rm diag}(1,1,-2)/\sqrt{3}$.  
  
The Bloch vector for the single-mode Hamiltonian is ${\bf H}_\omega$  
whereas the one for $\varrho_\omega$ is the polarization vector ${\bf  
P}_\omega$. Then the single-mode EOMs are  
\begin{equation}  
\dot{\bf P}_\omega={\bf H}_\omega\times {\bf P}_\omega\,,  
\end{equation}  
where the cross product is understood in the SU(3) sense: $({\bf  
A}\times {\bf B})_i=f_{ijk}A_jB_k$ where $i,j,k=1,\ldots,8$ and  
$f_{ijk}$ are the structure constants of the SU(3) Lie algebra.  
  
With the global polarization vector ${\bf P}=\int d\omega\,{\bf  
P}_\omega$ and ignoring the ordinary matter term, we write the  
single-mode Hamiltonian in the form  
\begin{equation}  
{\bf H}_\omega=\omega\,({\bf B}_{\rm H}+\alpha {\bf B}_{\rm L})  
+\mu\,{\bf P}\,.  
\end{equation}  
In the mass basis, the ``magnetic field'' components are  
\begin{equation}  
{\bf B}_{\rm H}=-\frac{2}{\sqrt3}\,{\bf e}_8  
\quad{\rm and}\quad  
{\bf B}_{\rm L}=-{\bf e}_3\,,  
\end{equation}  
representing the ``atmospheric'' and ``solar mass directions,''  
respectively.  
  
In the absence of ordinary matter, the EOM for the global  
polarization vector is  
\begin{equation}  
\dot {\bf P}=({\bf B}_{\rm H}+\alpha{\bf B}_{\rm L})  
\times{\bf M}\,,  
\end{equation}  
where the ``magnetic moment'' of the system is ${\bf M}=\int  
d\omega\,\omega\,{\bf P}_\omega$. In the mass basis this is  
\begin{equation}  
\dot {\bf P}=-\left(\frac{2}{\sqrt3}\,{\bf e}_8  
+\alpha\,{\bf e}_3\right)  
\times{\bf M}\,.  
\end{equation}  
The vector on the r.h.s.\ is orthogonal to both ${\bf e}_3$ and ${\bf  
e}_8$. The reason is that the matrices $\Lambda_3$ and $\Lambda_8$  
commute or, in other words, that $f_{a38}=0$ for $a=1,\ldots,8$ and  
the same for all permutations. As a consequence, the vector $\dot{\bf  
P}$ has no ${\bf e}_3$ or ${\bf e}_8$ component so that $\dot P_3=0$  
and $\dot P_8=0$. In a general basis this implies  
\begin{equation}  
d_t({\bf P}\cdot{\bf B}_{\rm H})=0  
\quad{\rm and}\quad  
d_t({\bf P}\cdot{\bf B}_{\rm L})=0\,.  
\end{equation}  
This is the equivalent of ``flavor-lepton number conservation''  
$d_t({\bf P}\cdot{\bf B})=0$ in the two-flavor  
context~\cite{Hannestad:2006nj, Raffelt:2007cb, Raffelt:2007xt}.  
In other words, in the three-flavor context we have two flavor-lepton  
numbers that are separately conserved. In the mass basis one  
concludes that  
\begin{eqnarray}\label{eq:polarcons}  
P_3&=&\varrho_{11}-\varrho_{22}\,,\nonumber\\  
P_8&=&\frac{\varrho_{11}+\varrho_{22}-2\varrho_{33}}{\sqrt{3}}  
\end{eqnarray}  
are conserved.  
  
\section{Synchronized MSW effect}                      \label{sec:MSW}  
  
As a first step in our analytic study we consider the matter-induced  
conversion of the initial $\nu_e$ flux as it passes the H and L  
level crossings, assuming that both lie deeply in the region where the  
neutrino-neutrino interaction is strong. As a result, the flavor  
oscillations are synchronized, meaning that all $\varrho_\omega$ stay  
pinned to each other. In other words, it is enough to study the  
evolution of the matrix of the total density $\varrho$. It obeys  
the~EOM  
\begin{equation}  
{\rm i}\partial_r\varrho(r)=[\Omega+\lambda(r){\sf L},\varrho(r)]\,,  
\end{equation}  
where in the mass basis  
\begin{equation}  
\Omega=\omega_0\,{\rm diag}\left(-{\textstyle\frac{1}{2}}\alpha,  
+{\textstyle\frac{1}{2}}\alpha,\pm 1\right)  
\end{equation}  
and $\omega_0=\langle \Delta m^2_{\rm atm}/2E\rangle$.  
  
For the Fermi-Dirac spectrum described in Sec.~\ref{sec:SNmodel} we  
find $\omega_0=0.710~{\rm km}^{-1}$. We will also use the notation  
\begin{eqnarray}  
\omega_{\rm H}&=&  
\left\langle\frac{\Delta m^2_{\rm atm}}{2E}\right\rangle  
=\omega_0=0.710~{\rm km}^{-1}\,,\nonumber\\  
\omega_{\rm L}&=&  
\left\langle\frac{\Delta m^2_{\odot}}{2E}\right\rangle  
=\alpha\omega_0=0.024~{\rm km}^{-1}\,.  
\end{eqnarray}  
The matrix of vacuum oscillation frequencies thus can also be written  
as $\Omega={\rm diag}\left(-{\textstyle\frac{1}{2}}\omega_{\rm L},  
+{\textstyle\frac{1}{2}}\omega_{\rm L},\pm \omega_{\rm H}\right)$.  
  
\begin{figure*}  
\includegraphics[angle=0,width=0.6\textwidth]{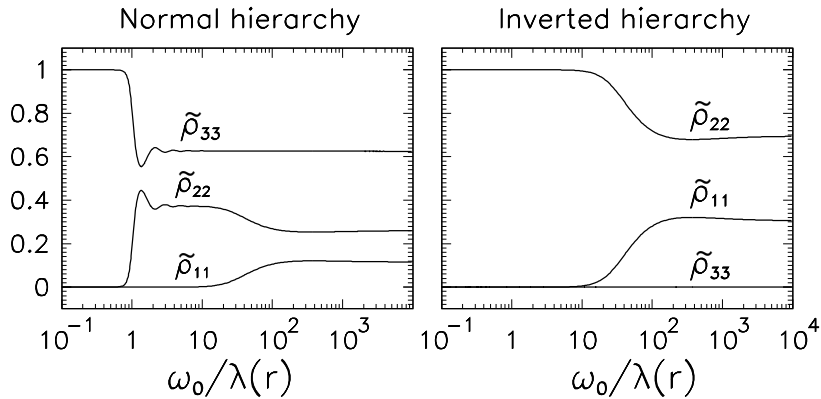}  
\caption{Evolution of the diagonal elements of the  matrix of  
densities $\tilde\varrho_{\omega}$ for a mode with  
$\omega=0.42$~km$^{-1}$, represented in the basis of instantaneous  
propagation eigenstates. Due to the synchronization, all modes behave  
similarly.\label{fig:MSW1}}  
\end{figure*}  
  
The assumed perfect synchronization implies that we can treat this  
system as an equivalent system with a single energy\footnote{An MSW transition in the  
presence of a dense neutrino gas was first treated  
in this way in the context of early-universe neutrino oscillations by  
Wong~\cite{Wong:2002fa}  
and by Abazajian, Beacom and Bell~\cite{Abazajian:2002qx}. They used  
the terminology ``collective MSW-like transformation'' and  
``synchronized MSW effect,'' respectively.}  
or rather with two  
fixed vacuum frequencies $\omega_{\rm H}$ and $\omega_{\rm L}$  
determining $\Omega$.  It is most useful to study its evolution in the  
basis of instantaneous propagation eigenstates where we denote the  
total matrix of density as $\tilde\varrho$. Since the only effect of  
the neutrino-neutrino interactions is to synchronize the oscillations  
and to reduce the system to an equivalent single-energy case, the  
propagation eigenstates are defined by the ordinary matter term for a  
monochromatic neutrino beam, wheras the neutrino-neutrino interaction  
plays no further role.  The propagation basis coincides with the weak  
interaction basis when the matter density is large so that our initial  
state is $\tilde\varrho={\rm diag}(0,0,1)$ in normal hierarchy, and  
${\rm diag}(0,1,0)$ in inverted hierarchy. If the subsequent evolution  
were perfectly adiabatic, the system would remain in this state so  
that the $\nu_e$ survival probability after the H and L crossings  
would be given by well known results~\cite{Dighe:1999bi}.  
  
We assume indeed that the evolution is adiabatic, except near the  
H and L  level crossings where in our system the jumping probabilities  
$P_{\rm H}$ and $P_{\rm L}$ need not be small. The neutrino mass-gap  
hierarchy ensures that the two crossings factorize with good  
approximation. In the normal hierarchy, the system encounters both  
crossings and the final occupation of the propagation eigenstates is  
given by the products of probabilities shown in Table~\ref{tab:MSW1}.  
In the inverted hierarchy, only the L crossing is encountered because  
the H level crossing is now in the antineutrino sector that is  
irrelevant in our case. Again, the final mass-state occupations are  
shown in Table~\ref{tab:MSW1}.  
  
\begin{table}  
\caption{Final occupations of the mass eigenstates after passing both  
MSW level crossings as described in the text.\label{tab:MSW1}}  
\begin{ruledtabular}  
\begin{tabular}{lllll}  
&\multicolumn{2}{l}{Normal hierarchy}  
&\multicolumn{2}{l}{Inverted hierarchy}\\  
\hline  
$\tilde\varrho_{11}$&$P_{\rm H}\,P_{\rm L}$&0.12&$P_{\rm L}$&0.31\\  
$\tilde\varrho_{22}$&$P_{\rm H}\,(1-P_{\rm L})$&0.26&$(1-P_{\rm L})$&0.69\\  
$\tilde\varrho_{33}$&$(1-P_{\rm H})$&0.62&$0$&0\\  
\end{tabular}  
\end{ruledtabular}  
\end{table}  
  
The jumping probability for incomplete adiabaticity is given to a  
good approximation by the so-called double-exponential  
formula~\cite{Petcov:1987zj, Krastev:1988ci, Fogli:2001pm}  
\begin{equation}  
P_{\rm H} = \frac{\exp(2 \pi R_{\rm H}\omega_{\rm H}  
\cos^2 \theta_{13})-1}  
{\exp(2 \pi R_{\rm H}\omega_{\rm H})-1}\,,  
\end{equation}  
where the scale height is  
\begin{equation}  
R_{\rm H} =  
\left|\frac{d \ln \lambda(r)}{dr}\right|^{-1}_{r=r_{\rm H}}\,.  
\end{equation}  
It has to be evaluated at the point of maximum violation of  
adiabaticity~\cite{Friedland:2000rn,Lisi:2000su,pmva-ricard}, given by  
$\omega_{\rm H}=\lambda(r_{\rm H})$~\cite{Fogli:2001pm}. The assumed  
power-law profile of Eq.~(\ref{eq:SNprofile2}) and our choices for  
$\omega_{\rm H}$ and $\theta_{13}$ imply $r_{\rm H}=1089$~km, $R_{\rm  
H}=21.8$~km, and $P_{\rm H}=0.38$. Analogous results pertain to  
$P_{\rm L}$ with ${\rm H}\to{\rm L}$ everywhere and the substitution  
$\theta_{13}\to\theta_{12}$. We find $r_{\rm L}=1166$~km, $R_{\rm  
L}=23.3$~km, and $P_{\rm L}=0.31$. These numerical results for  
$P_{\rm H}$ and $P_{\rm L}$ imply the numerical results for the final  
occupations shown in Table~\ref{tab:MSW1}.  
  
In the steep density profile used here, the H and L resonances  
  look spatially very close (Fig.~\ref{fig:profile}) so that one may  
  worry if they indeed factorize in the usual way. We stress that the  
  two resonances do not overlap, but one may still worry about  
  possible interference effects. However,  
we can compare the analytic results with a numerical three-flavor  
evolution, using the same power-law $\lambda(r)$ profile. In  
Fig.~\ref{fig:MSW1} we show the evolution of the diagonal elements of  
$\tilde\varrho_{\omega}$, for a mode with   $\omega=0.42$~km$^{-1}$,  
as a function of $\omega_0/\lambda(r)$. Due to the synchronization  
condition, all frequency modes behave the same.  
On this  
scale, the H crossing is at $\omega_0/\lambda(r)=1$ and the L  
crossing at $\omega_0/\lambda(r)=\alpha^{-1}=30$. The agreement  
between the numerical end states and the analytically predicted ones  
is striking.  
  
\begin{figure*}  
\includegraphics[angle=0,width=0.65\textwidth]{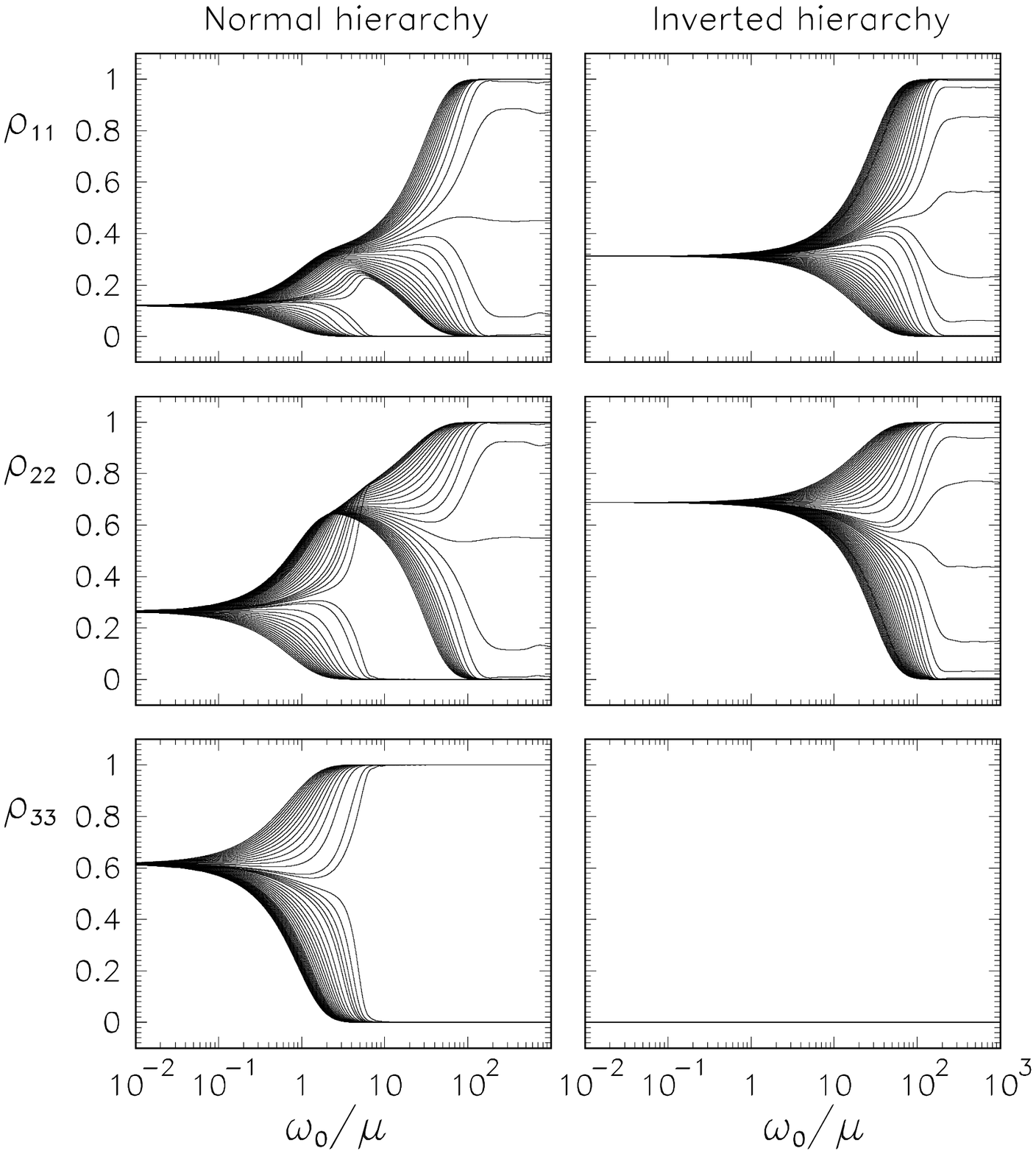}  
\caption{Evolution of the diagonal elements of $\varrho_\omega$ for  
our box-spectrum example. The mode density is increased around the  
splits.\label{fig:split1}}  
\vskip24pt  
\includegraphics[angle=0,width=0.65\textwidth]{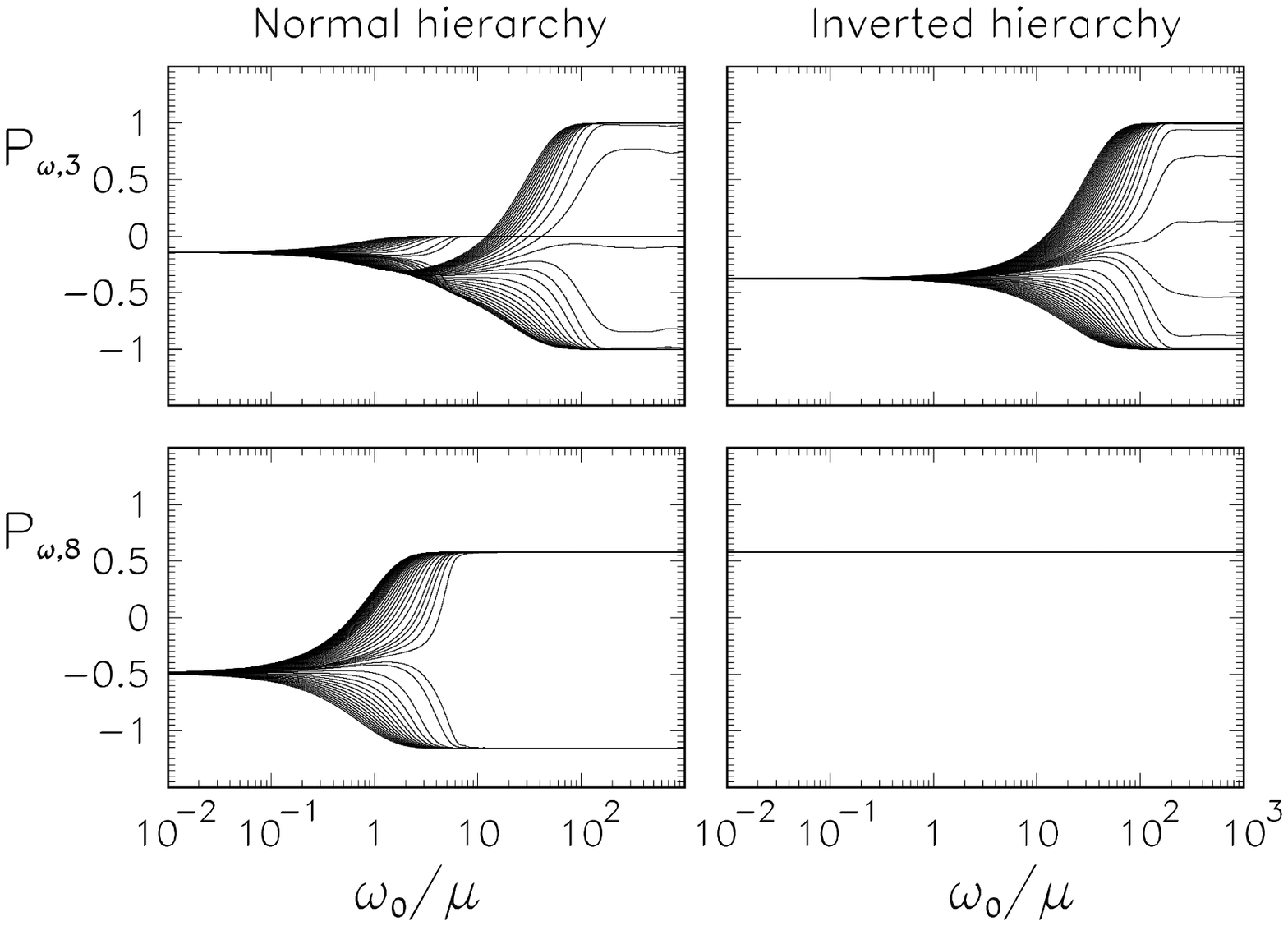}  
\caption{ 
Evolution of the 3 and 8  
components of ${\bf P}_\omega$ for our box-spectrum example. 
\label{fig:split2}}  
\end{figure*}

\section{Spectral split}                     \label{sec:splitting}  

After the system has passed the two MSW level crossings, the ordinary  
matter density quickly becomes negligible, whereas by assumption the  
neutrino-neutrino effects are still strong. The subsequent evolution  
to the point where the neutrino-neutrino interaction becomes  
negligible will then produce spectral splits in the same way as  
described in Refs.~\cite{Raffelt:2007cb, Raffelt:2007xt}. We can  
follow the previous two-flavor treatment almost step by step because  
the present three-flavor system is simplified by the mass-gap  
hierarchy $\alpha\approx 1/30\ll1$. While the two conserved  
flavor-lepton numbers present in the three flavor case lead to two  
spectral splits, these will occur in sequence and their dynamics  
factorizes in practice.  
  
The first split to develop is driven by the atmospheric mass  
difference and thus can be called the H split. As in  
Refs.~\cite{Raffelt:2007cb, Raffelt:2007xt} we go to a rotating  
frame, at first rotating around the ${\bf B}_{\rm H}$ direction. The  
single-mode Hamiltonians in this co-rotating frame are  
\begin{equation}  
{\bf H}_\omega \approx  
\left(\omega-\omega_{\rm H}^{\rm c}\right) {\bf B}_{\rm H}  
+\mu\,{\bf P}\,,  
\end{equation}  
neglecting for now the much smaller term
$\alpha \omega {\bf \rm B}_{\rm L}$.
This is justified  
because, when $\mu \agt \omega$ (and thus $\mu \gg \alpha\omega$),  
the ensemble of neutrinos is in a regime where we expect spectral  
splitting along ${\bf e}_8$ and synchronized oscillations along 
${\bf e}_3$. This factorization has been explicitly shown
in \cite{Dasgupta07}. 
Flavor conversion is thus driven primarily by ${\bf \rm  
B}_{\rm H}$, while ${\bf \rm B}_{\rm L}$ gives sub-leading  
corrections due to the synchronized oscillations. Similarly, when  
$\mu \sim \alpha\omega$, flavor conversion proceeds efficiently via a  
spectral split along ${\bf e}_3$ and is driven by ${\bf \rm B}_{\rm  
L}$, while ${\bf \rm B}_{\rm H}$ drives vacuum oscillations  
along~${\bf e}_8$.

Now, as $\mu$ adiabatically goes to zero, the co-rotation frequency  
$\omega_{\rm H}^{\rm c}$ approaches the final split frequency $\wsH$  
and the modes with $\omega>\wsH$ will orient themselves along ${\bf  
B}_{\rm H}$, those with $\omega<\wsH$ in the $-{\bf B}_{\rm H}$  
direction. The value of $\wsH$ is fixed by the conservation of $P_8$.  
Since the evolution associated with ${\bf B}_{\rm H}$ has saturated,  
we can next go into a frame rotating around ${\bf B}_{\rm L}$ where  
\begin{equation}  
{\bf H}_\omega \approx  
\left(\omega-\omega_{\rm L}^{\rm c}\right) {\bf B}_{\rm L}  
+\mu\,{\bf P}  
\end{equation}  
and repeat the analogous argument.  
  
To illustrate the  dynamics of the split in a form similar to  
Refs.~\cite{Raffelt:2007cb, Raffelt:2007xt} we consider an explicit  
example with an initial ``box spectrum'' at high density of the form  
\begin{equation}  
\varrho_{ee}(\omega)=  
\cases{(2\omega_0)^{-1}&for $0\leq\omega\leq2\omega_0$\,,\cr  
0&otherwise\,.\cr}  
\end{equation}  
At high densities $\rho_{ee}(\omega)$ coincides with  
${\tilde \rho}_{33}(\omega)$ in normal hierarchy and with  
 ${\tilde \rho}_{22}(\omega)$ in inverted hierarchy.  
After the MSW crossings the spectrum is still of box shape because  
of the assumed strong neutrino-neutrino interaction, but now has the  
$\varrho_{11}$, $\varrho_{22}$, and $\varrho_{33}$ components shown  
in Table~\ref{tab:MSW1}. Note that after the MSW transitions we  
neglect ordinary matter so that the propagation eigenstates are  
identical with the mass eigenstates and $\tilde\varrho=\varrho$.  
  
\subsection{Normal hierarchy}  
  
At first we study the evolution caused by the neutrino-neutrino  
interactions numerically. To this end we use the usual $\mu(r)\propto  
r^{-4}$ profile of Eq.~(\ref{eq:muprofile}) with the coefficient  
$\mu_0$ of Eq.~(\ref{eq:mucoefficient}). In this way the evolution is  
strongly but not perfectly adiabatic. Of course, the analytic  
results, based on the conservation of flavor-lepton number, apply in  
the perfectly adiabatic limit which here requires $\mu_0\to\infty$.  
  
In Fig.~\ref{fig:split1} (left column) we show the evolution in the  
mass basis of the diagonal elements of $\rho_\omega$ for 50~modes as  
a function of $\omega_0/\mu$. All modes start with the same initial  
condition prepared by the MSW transitions. Around $\omega_0/\mu=1$  
one recognizes a first H split that affects all components. Around  
$\omega_0/\mu=\alpha^{-1}$ one recognizes a second split, the L  
split, that affects only $\rho_{11}$ and $\rho_{22}$. During the H  
split, the modes with $\omega<\wsH$ tend to $\rho_{33}\to 0$, while  
those with $\omega > \wsH$ approach $\rho_{33}\to 1$.  Then  
$\rho_{11}$ and $\rho_{22}$ modes with $\omega  
>\wsH$ go to 0, as implied by the conservation of the trace of $\rho$.  
While the modes with $\omega < \wsH$ rise towards higher values of  
$\rho_{11}$ and $\rho_{22}$, they encounter the L split at a  
frequency $\wsL<\wsH$.  At this split, for $\omega<\wsL$, the  
$\rho_{11}$ approach 1 and the $\rho_{22}$ approach 0, and vice versa  
for $\wsL<\omega<\wsH$. As a result of imperfect adiabaticity some  
modes do not reach these extreme values, but get frozen earlier.  
  
In Fig.~\ref{fig:split2} (left column) we show the same case in terms  
of the mass-basis $P_{\omega,3}$ and $P_{\omega,8}$ components. We  
observe that in $P_{\omega,8}$ only the H split operates whereas in  
$P_{\omega,3}$ the H and L splits operate in sequence.  
  
\begin{figure*}  
\includegraphics[angle=0,width=0.65\textwidth]{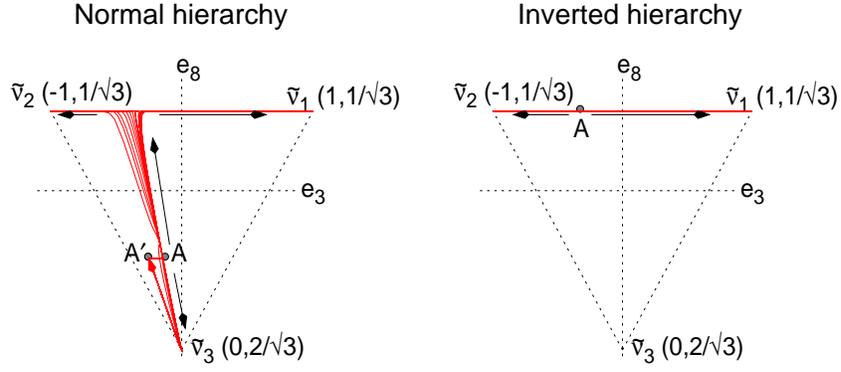}  
\caption{\label{fig:tplot}Projection of the polarization vectors  
${\bf  
    P}_{\omega}$ on the ${\bf e}_{3}$-${\bf e}_8$ plane for our box-example. 
The vertices of the triangle represent pure (instantaneous) mass eigenstates. 
After both MSW transitions, the system is at the point A in the interior of 
the triangle. (See the text for details.)  
\label{fig:triang}}  
\end{figure*}  
  
The situation can be visualized in terms of the ${\bf e}_3$--${\bf  
  e}_8$ triangle diagram~\cite{Dasgupta07} shown in Fig.~\ref{fig:triang}.  
 Each point in the  
interior and on the boundary of the triangle represents the projection  
of the polarization vector ${\bf P}_\omega$ in the ${\bf e}_3$--${\bf e}_8$ plane.  
Neutrinos from the $\nu_e$ burst start in the state $\nu_e \approx  
\tilde{\nu}_3$, where by ``tilde'', we represent the instantaneous  
mass eigenstates.  The H crossing shifts the neutrino state from the  
$\tilde{\nu}_3$ vertex towards the $\tilde{\nu}_2$ state, but only  
partially, due to the semiadiabatic nature of the transition.  
After that crossing, all neutrinos find themselves at the point A$^{\prime}$  
inside the triangle.  The L  
crossing further transports the state along a line parallel to the  
$\tilde{\nu}_2$--$\tilde{\nu}_1$ edge towards $\tilde{\nu}_1$, again  
only partly due to the semiadiabaticity.  Before the split, all the  
neutrinos are thus at a point A in the interior of the triangle.  
  
The H split takes the $\omega > \wsH$ modes towards the  
$\tilde{\nu}_3$ state ($P_{\omega,3}=0, P_{\omega,8}=-2/\sqrt{3}$)  
and the modes $\omega < \wsH$ towards some combination of  
$\tilde{\nu}_1$ and $\tilde{\nu}_2$, while conserving the  
total $P_3$ and $P_8$.  
Since $\alpha \ll 1$, the H and L splits are well separated and the  
high-$\omega$ modes reach the $\tilde{\nu}_3$ vertex, i.e. the  
H split saturates, before the L split begins.  
The low-$\omega$ modes propagating towards the  
$P_{\omega,8}=1/\sqrt{3}$ line encounter the L split that tends to  
take the $\omega > \wsL$ modes towards  
$\tilde{\nu}_2$ ($P_{\omega,3}=-1, P_{\omega,8}=1/\sqrt{3}$)  
and the $\omega < \wsL$ modes towards  
$\tilde{\nu}_1$ ($P_{\omega,3}=1, P_{\omega,8}=1/\sqrt{3}$).  
In the adiabatic limit, given sufficient time to propagate,  
the H and L splits result in all neutrinos reaching one of  
the three vertices of the ${\bf e}_3$--${\bf e}_8$ triangle.

Using the conservation law for $P_3$ and $P_8$ of  
Eq.~(\ref{eq:polarcons}) one can evaluate the  split frequencies  
$\wsH$ and $\wsL$. For $\omega<\wsH$ we have $P_{\omega,8}\to  
1/{\sqrt 3}$, while for $\omega>\wsH$ they reach $-2/{\sqrt 3}$. In  
the limit of perfect adiabaticity, the conservation of $P_8$ implies  
\begin{equation}  
2\omega_0 P_{\omega,8}^{0}=\frac{1}{\sqrt3}\,\wsH  
- \frac{2}{\sqrt3}\,\left(2\omega_0 - \wsH\right)\,,  
\end{equation}  
where $P_{\omega,8}^{0}$ is the common value of $P_{\omega,8}$  
before the split begins.  
In our example, $P_{\omega,8}^{0}=-0.50$, leading to  
$\wsH=0.76\,\omega_0$.  
  
\begin{figure*}  
\includegraphics[angle=0,width=0.65\textwidth]{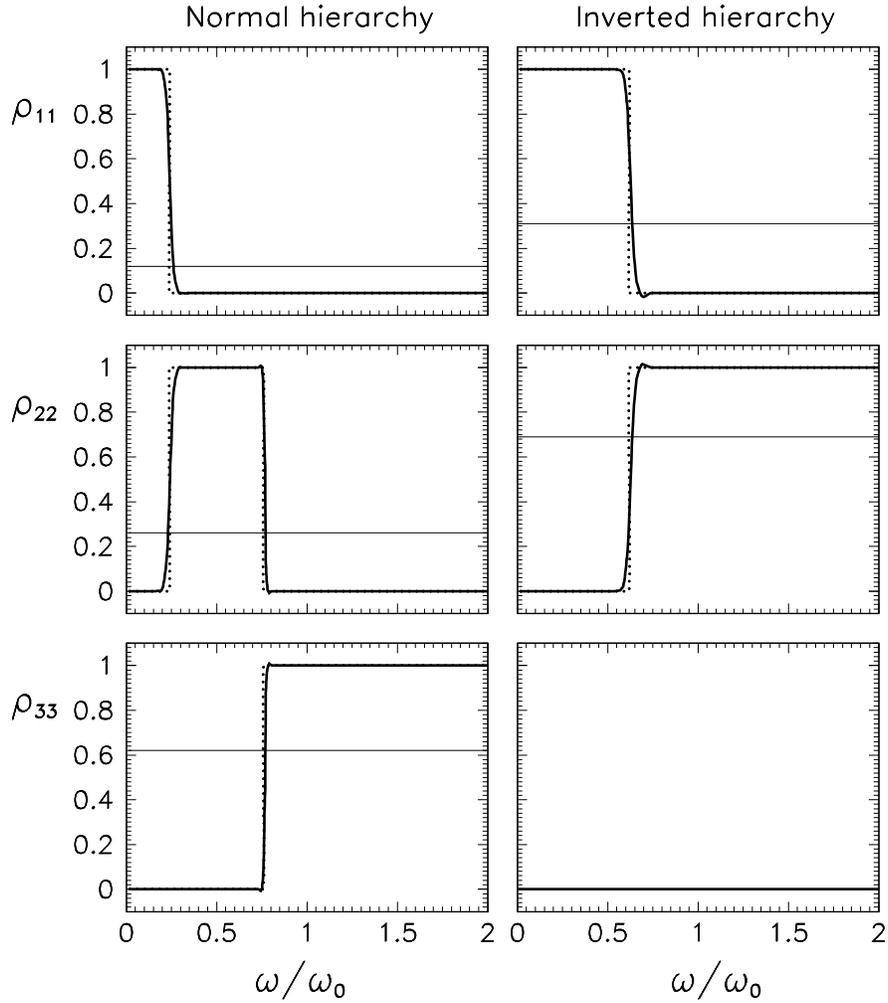}  
\caption{Diagonal elements of $\varrho_\omega$.  
Thin line: initial box spectrum. Thick line: final numerical spectrum.  
Dotted line: final spectrum in the adiabatic limit.  
\label{fig:spec1}}  
\end{figure*}  
  
  
\begin{figure*}  
\includegraphics[angle=0,width=0.65\textwidth]{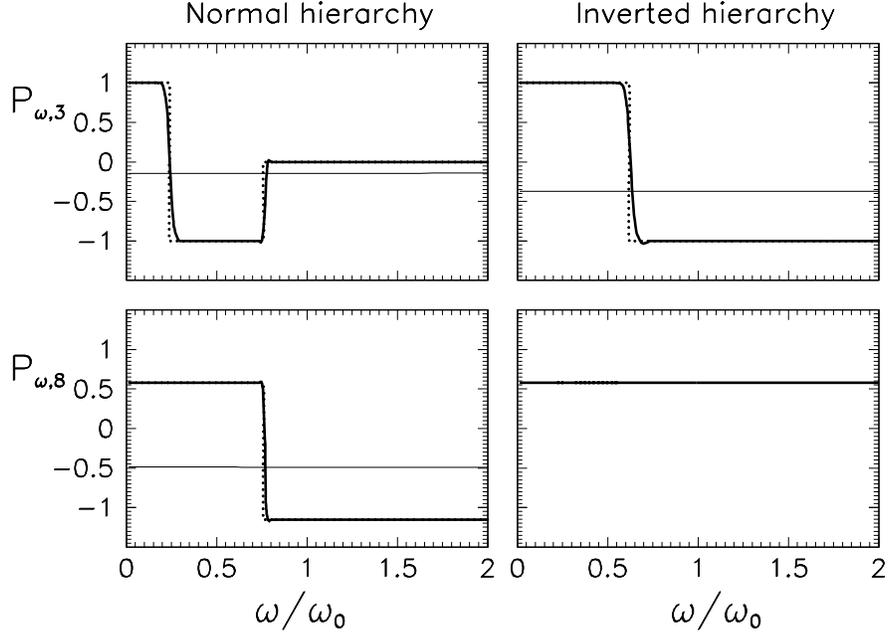}  
\caption{
The 3 and 8 components of ${\bf P}_\omega$. Convention for lines 
is same as in Fig.~\ref{fig:spec1}.
\label{fig:spec2}}  
\end{figure*}

\begin{figure*}  
\includegraphics[angle=0,width=0.60\textwidth]{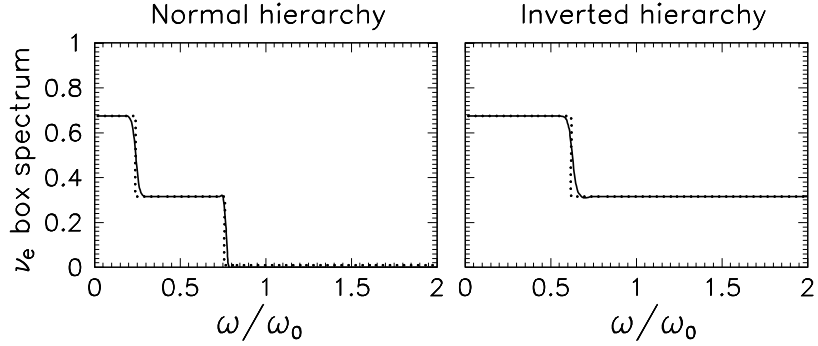}  
\caption{
The $\nu_e$ component at a large distance so that the 
different $\omega$ modes have kinematically decohered.
Convention for lines is same as in Fig.~\ref{fig:spec1}.
\label{fig:spec3}}  
\end{figure*}  
  
\begin{figure*}  
\includegraphics[angle=0,width=0.60\textwidth]{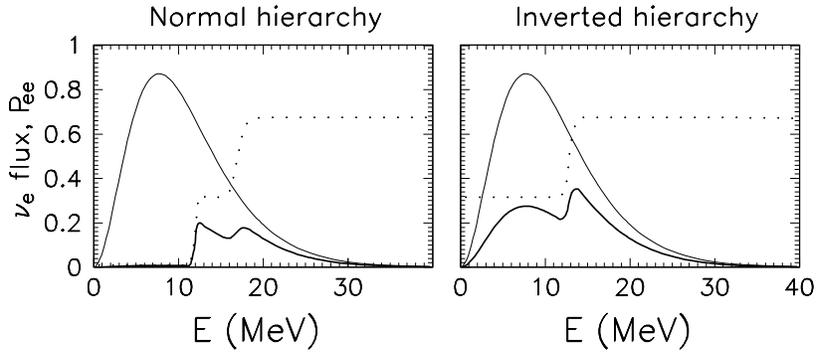}  
\caption{Initial (thin) and final (thick) spectrum for a Fermi-Dirac distribution  
with the parameters described in Sec.~\ref{sec:SNmodel}. The  
numerical final spectrum is for our toy-model SN where the MSW  
crossings and spectral-split region are far  
separated. Dotted curves represent the survival probability $P_{ee}$  
for electron neutrinos.\label{fig:spec4}}  
\end{figure*}

When the H split saturates, all modes with $\omega > \wsH$ have  
$P_{\omega,8} = -2/\sqrt{3}$, and hence $P_{\omega,3} = 0$ due to  
the conservation of the norm of ${\bf P}$. These modes have reached  
the bottom vertex of the ${\bf e}_3$--${\bf e}_8$ triangle and hence  
cannot split further due to the L split. On the other hand, for  
modes with $\omega<\wsH$ a second split in $P_{\omega,3}$ happens.  
These modes approach $P_{\omega,3}=+1$ for $\omega < \wsL$ and  
$P_{\omega,3}=-1$ for $\omega>\wsL$. Applying the conservation law  
for $P_3$ gives~us  
\begin{equation}  
2 \omega_0 P_{\omega,3}^{0}= \wsL - \left(\wsH-\wsL\right)\,.  
\end{equation}  
In our example $P_{\omega,3}^{0}=-0.14$ so that $\wsL=0.24\,\omega_0$.

We show in Figs.~\ref{fig:spec1} and~\ref{fig:spec2} the mass-basis  
spectra of the diagonal elements of $\varrho$ and of $P_3$ and $P_8$.  
Thin lines are the MSW-prepared initial spectra. Thick lines show the  
numerical end states, corresponding to the split diagrams of  
Figs.~\ref{fig:split1} and~\ref{fig:split2}. Dotted lines show the  
adiabatic limiting behavior based on the lepton-number conservation  
laws. Once more the agreement is striking. Imperfect adiabaticity  
leads to a smoothening of the splits which otherwise are sharp  
spectral steps.  
   
Finally, in Fig.~\ref{fig:spec3} we show the $\nu_e$ spectrum after  
the two splits. The solid curve is the numerical result, where  
the vacuum oscillations between the SN and the observer have  
been averaged (kinematical decoherence between different $\omega$  
modes), i.e.,  
\begin{equation}  
\varrho_{ee} = U_{e1}^2 \varrho_{11} + U_{e2}^2 \varrho_{22} +  
U_{e3}^2 \rho_{33} \,\ .  
\label{eq:aver}  
\end{equation}  
The dotted curve is our analytic result in the adiabatic limit. This  
result is easily explained if we observe that in Eq.~(\ref{eq:aver})  
in the case of maximal 23 mixing one has $U_{e1}^2 =    \cos^2 \theta_{13} \cos^2  
\theta_{12}$, $U_{e2}^2 = \cos^2 \theta_{13} \sin^2 \theta_{12}$, and $U_{e3}^2 =  
 \sin^2 \theta_{13}$. Therefore,  
\begin{equation}\label{eq:cases1}  
\rho_{ee}\simeq\cases{  
 \cos^2\theta_{12} &for $\omega < \wsL\,,$\cr  
 \sin^2\theta_{12} &for $\wsL < \omega < \wsH\,,$\cr  
 \sin^2\theta_{13} &for $\wsH<\omega\,.$\cr}  
\end{equation}  
Again the agreement between the analytic and numerical results is  
very good.

\subsection{Inverted hierarchy}  
  
For the inverted hierarchy we show the analogous information in the  
right-handed columns of Figs.~\ref{fig:split1}--\ref{fig:spec3}. The  
initial state here is $\tilde{\nu}_2$. The nonadiabatic L crossing  
takes the neutrino states partly towards $\tilde{\nu}_1$. After the  
L crossing and before the split, the neutrino state for all modes is  
along the $\tilde{\nu}_1$--$\tilde{\nu}_2$ edge, at A as shown in  
Fig.~\ref{fig:triang} (right column), where $P_{\omega,8} =  
1/\sqrt{3}$. Since all neutrinos already are in one of the extreme  
values of $P_{\omega,8}$, the H split is inoperational. This  
corresponds to $\rho_{33}$ remaining in its MSW-prepared initial  
value of~0. The L split takes $\rho_{11} \to +1$ for $\omega <\wsL$  
and $\rho_{11} \to 0$ for $\omega>\wsL$, and vice versa for  
$\rho_{22}$. In the inverted hierarchy we have an effective  
two-flavor case in the $\nu_1$-$\nu_2$ subsector. This is a consequence of the  
MSW-prepared initial condition. Initially  
$P_{\omega,8}^{0}=1/\sqrt3$. Applying now the conservation of  
$P_{8}$ we obtain $\wsH=2\omega_0$, i.e., the split occurs at the  
edge of the box and thus is not visible. The conservation law for  
$P_3$ and using in our case $P_{\omega,3}^{0}=-0.38$, one obtains  
$\wsL=0.62\,\omega_0$. For the electron flavor we predict for the  
final spectrum  
\begin{equation}\label{eq:cases2}  
\rho_{ee} \simeq \cases{  
 \cos^2\theta_{12} &for $\omega < \wsL\,,$\cr  
 \sin^2\theta_{12} &for $\wsL<\omega\,,$\cr}  
\end{equation}  
in agreement with the numerical result.

Note that in the inverted hierarchy, there is only one split.  
This is because $P_{\omega,8}$ is already at an extreme value  
before the split can begin.  
In general, there are two splits if the neutrino state before the  
split is in the interior of the ${\bf e}_3$--${\bf e}_8$ triangle,  
one split if it is along one of the edges of the triangle (as in this  
case), and no split occurs if the neutrino state is at any of the  
three  vertices (See Fig.~\ref{fig:triang}).

\subsection{Fermi-Dirac spectrum}  
  
It is straightforward to extend these arguments to a general  
spectrum, e.g.\ a Fermi-Dirac spectrum $f(\omega)$.  
The conservation laws imply for the  
normal hierarchy  
\begin{eqnarray}  
 \sqrt{3} P_{\omega,8}^{0} &=&  
\int_{0}^{\wsH}\,d\omega\,f(\omega)\  
-2\int_{\wsH}^{\infty}\,d\omega\,f(\omega)\,,  
\nonumber\\  
P_{\omega,3}^{0} &=&  
\int_{0}^{\wsL}\,d\omega\,f(\omega)\  
-\int_{\wsL}^{\wsH}\,d\omega\,f(\omega)\,.  
\end{eqnarray}  
On the other hand, for the inverted hierarchy we find  
\begin{eqnarray}  
 P_{\omega,3}^{0}  &=&  
\int_{0}^{\wsL}\,d\omega\,f(\omega)\  
-\int_{\wsL}^{\infty}\,d\omega\,f(\omega)\,.  
\end{eqnarray}  
These relations allow us to calculate $\wsH$ and $\wsL$. Note that  
these results are exact only in the limit of an infinite mass-gap  
hierarchy, i.e., for $\alpha\to 0$, where the H and L splits  
perfectly factorize.  
Once the split frequencies have been found, the corresponding  
energies are $E_{\rm H}^{\rm s}=\Delta m^2_{\rm atm}/2\wsH$ and  
$E_{\rm L}^{\rm s}=\Delta m^2_{\odot}/2\wsL$.  
  
For our schematic SN model where the MSW level crossings and the  
spectral split region are widely separated, we find $E_{\rm H}^{\rm  
s}=11.9$~MeV and $E_{\rm L}^{\rm s}=16.9$~MeV in the normal  
hierarchy, and $E_{\rm L}^{\rm s}=12.7$~MeV in the inverted  
hierarchy. We show the initial and final $\nu_e$ spectra for the  
Fermi-Dirac case in Fig.~\ref{fig:spec4}. Once more, we have  
coarse-grained over neighboring modes, representing the effect of  
kinematical decoherence.   
  
\section{Conclusions}                          \label{sec:conclusions}  
  
We have studied the three-flavor evolution of a $\nu_e$ burst that  
first undergoes two MSW level crossings driven by $\Delta m^2_{\rm  
atm}$ and $\Delta m^2_\odot$, respectively, and then undergoes  
spectral splits by the adiabatically decreasing strength of the  
neutrino-neutrino interaction. This case study of an MSW-prepared  
spectral split serves as a proxy for the recent numerical study of  
the prompt $\nu_e$ burst in an O-Ne-Mg core collapse SN. Here, the  
matter density profile is so steep that the sequence between MSW  
crossings and collective neutrino oscillations is reversed from what  
would be expected in a traditional iron-core SN.  
  
First we have analytically estimated the population of the propagation  
eigenstates after the MSW transformations.  Due to the sharply falling  
matter density at the edge of the core of the star, the MSW transitions  
are not completely adiabatic, which helps in satisfying a precondition  
for the spectral splits to take place.  We have used the  
well-known analytic double-exponential formula for calculating the  
jump probabilities.  
  
We have studied numerically the subsequent spectral splits. We have  
analytically determined the split frequencies on the basis of two  
conserved flavor-lepton number combinations that supersede the  
single conservation law encountered in a two-flavor situation. The  
neutrino mass-gap hierarchy allows for a factorization of an H split  
and an L split, similar to the factorization of the MSW effect into  
an H and an L crossing. The  dynamics of the split evolution is  
clearly seen to be a two-step process.  
  
The dynamics of the two spectral splits can be understood in terms of the  
motion of the neutrino state in the ${\bf e}_3$--${\bf e}_8$  
triangle diagram, which can explain many of the features of  
neutrino evolution qualitatively.  
The number of possible splits can be deduced by the location of the  
neutrino state inside the triangle.  
We have also shown how the positions  
of the splits can be calculated accurately given the initial  
neutrino spectra, and calculated the $\nu_e$ survival probability  
analytically, that matches the numerical computations.  
  
Our analytic treatment accounts very nicely for the numerical  
findings of Duan et al.~\cite{Duan:2007sh}. In their case the MSW  
conversion and the spectral splits are spatially very close so that  
it is not a priori obvious that our schematic model would be a good  
representation. In our treatment we have enforced a clear separation  
between the MSW region and the spectral-split region by assuming the  
limit of large neutrino-neutrino interactions, a limit that also  
ensures that the MSW transition is perfectly synchronized and that  
the spectral split is perfectly adiabatic. A posteriori, however, our  
interpretation as a MSW prepared spectral split appears nicely  
justified and quantitatively appropriate. Our treatment is also a  
useful application of the three-flavor oscillation machinery  
developed by two of us recently~\cite{Dasgupta07}.  
  
It appears that the impact of collective neutrino oscillations on  
the propagation of the prompt $\nu_e$ burst is conceptually and  
quantitatively well under control. Characteristic signatures of  
these flavor transitions in large underground detectors have also  
been recently investigated~\cite{Lunardini:2007vn}. What remains is  
to observe these features in the neutrino signal of the next  
galactic~SN.

\section*{Note Added}  
  
After our manuscript was completed, a paper by Duan, Fuller and Qian  
appeared that treats three-flavor split  
phenomena~\cite{Duan:2008za}. The results partly overlap with our  
work.  
  
  
\begin{acknowledgments}  
In Munich, this work was partly supported by the Deutsche  
Forschungsgemeinschaft (grant TR-27 ``Neutrinos and Beyond''), by the  
Cluster of Excellence ``Origin and Structure of the Universe'' and by  
the European Union (contract No.\ RII3-CT-2004-506222). 
In Mumbai,  
partial support by a Max Planck India Partnergroup grant is  
acknowledged. 
B.D. and A.D. also thank IMSc, Chennai, for their hospitality
during WHEPP-X, when this work was completed.
A.M.\ acknowledges support by the Alexander von Humboldt  
Foundation.  A.M. acknowledges kind hospitality at 
the Tata Institute for Fundamental Research during 
the development of this work.  
\end{acknowledgments}  
  

\end{document}